# Rogue events in spatio-temporal numerical simulations of unidirectional waves in basins of different depth

A. Slunyaev[1,2)], A. Sergeeva[1,2)], I. Didenkulova[1,2,3)]

1) Institute of Applied Physics, Nizhny Novgorod, Russia, Slunyaev@hydro.appl.sci-nnov.ru
2) Nizhny Novgorod State Technical University, Nizhny Novgorod, Russia
3) Marine Systems Institute at Tallinn University of Technology, Tallinn, Estonia

**Abstract**

The evolution of unidirectional nonlinear sea surface waves is calculated numerically by means of solutions of the Euler equations. The wave dynamics corresponds to quasi-equilibrium states characterized by JONSWAP spectra. The spatio-temporal data are collected and processed providing information about the wave height probability and typical appearance of abnormally high waves (rogue waves). The waves are considered at different water depths ranging from deep to relatively shallow cases ($k_p h > 0.8$, where $k_p$ is the peak wavenumber, and $h$ is the local depth). The asymmetry between front and rear rogue wave slopes is identified; it becomes apparent for sufficiently high waves in rough sea states at all considered depths. The lifetimes of rogue events may reach up to 30-60 wave periods depending on the water depth. The maximum observed wave has height of about 3 significant wave heights. A few randomly chosen in-situ time series from the Baltic Sea are in agreement with the general picture of the numerical simulations.

## 1. Introduction

Direct numerical simulations of irregular sea waves have attracted much interest due to at least the following reasons. The estimation of probability and significance of extreme wave events needs big data of waves, which is not currently available from merely in-situ measurements. One of the most representative statistics of sea waves discussed by Christou & Ewans (2014) contains 122 million individual waves, what would correspond to less than 40 years of continuous measurements in a single point, assuming all waves have period 10 sec. It is also important to keep in mind that if the statistical averaging is performed inaccurately (e.g. different sea states should not be mixed), the result will be meaningless and therefore the available statistics diminishes significantly. Meanwhile existing regulations for permissible wave loads on offshore platforms operate with wave parameters, which may be observed once in 100 or even 10 000 years. The significance of the problem has received the general recognition in recent years. The phenomenon of rogue waves assumes that extreme waves are in fact much more frequent, than it has been usually assumed, due to various physical mechanisms, see reviews Kharif & Pelinovsky (2003), Dysthe et al (2008), Kharif et al (2009), Slunyaev et al (2011) and references therein.

Controlled experiments may solve the problem of statistical homogeneity of the wave data, though laboratory measurements are time and money consuming. The numerical simulations of kinetic equations cannot serve as a solution of the rogue wave problem, since the most interesting situations correspond to essentially nonlinear wave dynamics, when the assumption of near-Gaussianity of the waves is broken. The direct numerical simulation of large wave ensembles within primitive water equations with the purpose to produce rich statistics has become recently a feasible task due to the new fast algorithms and availability of powerful computers. The statistical data obtained from the direct numerical simulations have potential to fill the gap in trustworthy in-situ data. As a result, a number of papers on intensive stochastic simulations of approximate and full equations appeared in 2000-s (including, among others, Onorato et al. 2001, 2002, 2009, Dysthe et al. 2003; Janssen 2003; Socquet-Juglard et al. 2005; Chalikov 2005, 2009; Ducrozet et al. 2007; Xiao et al, 2013, see also a brief review in Sergeeva & Slunyaev 2013). The general approach is as follows: the initial condition is specified in the form of a realization of irregular waves with given spectrum and random phases. The evolution of the wave ensembles is simulated directly (most often – for not more than a few tens of wave periods); the wave data at a few subsequent instants compose the statistical data. The validity of numerical simulations has been confirmed in many comparative laboratory experiments (e.g. Onorato et al. 2009, Shemer et al. 2010). Most of these studies concerned the deep water regime, at the same time there are numerous observations of anomalously high waves in the coastal area (Didenkulova et al. 2006, 2013; Nikolkina & Didenkulova 2012).

Several main outcomes of the deep water wave research may be formulated. First, if the initial condition is characterized by sufficiently narrow frequency and angle spectra (for a given wave energy), then waves undergo an extreme transient state, when the kurtosis and probability of high waves grow significantly; the duration of this transient process is 1–2 characteristic scales of the cubic nonlinearity (see Slunyaev 2010). The process, which is responsible for the extreme state, is the modulational (Benjamin – Feir) instability; the instability criterion (Benjamin – Feir Index, BFI) was suggested to assess the significance of this effect for a given spectrum (Onorato et al. 2001; Janssen 2003). BFI is often considered to be a warning criterion of high likelihood of occurrence of abnormally high waves (rogue waves), when $BFI > 1$. After the stage of the modulational instability the portion of high waves decreases, and momentary BFI drops down to $BFI \sim 1$. Stationary sea states seem to be modulationally stable. Thus, extreme wave statistics due to the modulational instability takes place in transient states; the effects of modulational instability are much less pronounced in stationary sea conditions (see discussion in Slunyaev & Sergeeva 2011; Slunyaev et al. 2015). In particular, the relation between the BFI index and the wave statistics is not confirmed in fully nonlinear simulations of unidirectional waves by Chalikov (2009).

Besides simulations of model spectra, the direct numerical simulations are used to reconstruct the sea condition on the basis of hindcasting data (Bitner-Gregersen et al. 2014; Dias et al. 2015).

The majority of the studies concern the infinitively deep water case. Shallow water simulations were performed by Pelinovsky & Sergeeva (2006), which claim that the probability of high waves increases when the shallow water nonlinear parameter, the Ursell number, is large. Papers by Sergeeva et al. (2011, 2014) and Trulsen & Zeng (2012), Zeng & Trulsen (2012), Viotti & Dias (2014) consider variable depth and thus correspond to the situation, when waves are not in a stationary state. The nonlinear mechanism of adiabatic wave enhancement due to decreasing water depth is discussed in Slunyaev et al. (2015). Very recent stochastic simulations of waves in a constant-depth basin are reported in Fernandez et al. (2016); there the numerical simulations are compared with laboratory measurements as well. Since the modulational instability becomes three-dimensional for $kh < 1.36$ (where $k$ is the carrier wavenumber and $h$ is the local depth), strictly speaking the 3D simulations should be carried out to take into account the effect of modulational instability in such shallow water (e.g. Fernandez et al. 2014). The longitudinal modulational instability vanishes when $kh < 1.36$, thus the high wave probability is expected to decrease in shallow water. Indeed, the calculated probability of large wave heights was below the Reyleigh distribution in the in-situ measurements by Mori et al. (2002), Didenkulova and Anderson (2010), Mai et al. (2010),

Didenkulova (2011); all that measurements were performed in relatively shallow conditions. Numerical simulations of the Euler equations for relatively short distances performed by Fernandez et al. (2016) confirm this conclusion. In Gemmrich & Garrett (2010) the dynamical nonlinear effects were disregarded, and only second non-resonant harmonic was taken into account. Their simulations showed good agreement with in-situ data retrieved at deep and relatively shallow water conditions

Besides wave height probability functions, there are other important parameters characterizing abnormally high waves (rogue waves), such as the characteristic life time, wave shapes etc. The 'holes in the sea', i.e., very deep wave troughs are considered sometimes to be even more dangerous than waves with high crests. The numerical simulations of deep water waves within the framework of the Euler equations reported in Sergeeva & Slunyaev (2013) and Xiao et al. (2013), showed specific asymmetry of strongly nonlinear rogue waves: the rear wave slope is usually higher than the preceding front slope. At the same time, the well known examples of rogue waves on Agulhas current have different shape – a very long and deep preceding trough, which ends with a high crest (Mallory 1974; Lavrenov 2003). The lifetimes of rogue events, observed in Sergeeva & Slunyaev (2013), were up to 60 wave periods, if the definition of a rogue wave

$$\frac{H}{4\sigma} > 2, \qquad (1)$$

may fail for short times. Here $H$ is the wave height and $\sigma$ is the root-mean-square surface displacement. The significant wave height is roughly equal to $4\sigma$. Unexpectedly long lifetimes of rogue waves were also mentioned by other researchers.

In this paper we generalize the study of irregular sea waves, reported in Sergeeva et al. (2013), to conditions of an intermediate depth. The wave system is assumed to stand in a quasi-stationary condition with an exception of small energy leakage due to effects associated with occasional wave breaking. The ability of the approach to reproduce the given wave condition in the numerical simulation is examined. The effect of a finite depth on the wave statistics and appearance of rogue waves is in the focus of the study.

In Section 2 the approach for the numerical simulation is described. In Section 3 we define the simulated wave conditions. The wave height probability is discussed in Section 4, while in Section 5 the asymmetry and lifetimes of rogue waves are addressed. In Section 6 a few examples of in-situ wave records from the Black Sea are considered in view of the results of the numerical simulations. The main conclusions are formulated in the end.

## 2. Approach for the simulation of stationary sea states

The approach for numerical simulations generally follows the one used in Sergeeva & Slunyaev (2013). The wave data results from numerical simulations of the potential Euler equations by means of the High Order Spectral Method (West et al. 1987). Parameter $M$ characterizes the nonlinear property of the method; the nonlinear surface boundary conditions are solved accurate to order $M$ of the wave steepness (i.e., up to $M + 1$ wave-wave interactions are resolved).

Each wave realization is taken in a form of a linear superposition of waves with a given wavenumber spectrum and random phases (i.e., the assumption of a Gaussian wave field holds); it is used as the initial condition for a simulation. The surface elevation and surface velocity potential are related according to the linear relation for waves at a constant depth $h$. During the initial stage, which lasts for about 20 wave periods, the nonlinear terms in the solver are being enabled slowly; this transition procedure helps to adjust the waves to nonlinearity (Dommermuth 2000), and eventually to reduce the level of spurious waves. During this period the energy of the wave system may change, thus the values of $\sigma$ in Table 1 are slightly different from values which characterize the initial condition. The averaged parameters which characterize the peak period and wavenumber, and wave intensity are given in Table 1.

For the initial condition we use the JONSWAP spectrum with peakedness $\gamma = 3$ for less intense waves, and $\gamma = 3.3$ for steeper waves; these parameters are similar to ones used in Sergeeva & Slunyaev (2013); the peak wave period of the initial condition is 10–10.5 s. Each wave realization has length about 10 km; it is simulated for 20 min in a domain with periodic boundary conditions. The output data grid consists of 2048 × 2048 points, what corresponds to about 5 m × 0.6 s mesh size. One hundred wave realizations are used for each condition, what gives in total $O(10^7)$ single waves in time series or in space series for each case. A few intermediate depths are concerned, see Table 1. The infinitively deep condition was simulated and discussed in Sergeeva & Slunyaev (2013), they correspond to the shaded columns in Table 1.

## 3. Simulated sea states

The simulated series correspond to two intensities of the initial conditions; series A corresponds to less intense waves, which mainly do not break (only 2 realizations from series $A_{0.8}$ caused wave breaking and were redone). Realizations from series A are simulated with a high order of the code nonlinearity, $M = 6$, which guarantees almost fully nonlinear simulations of the Euler equations, with very small energy change (relative error is of the order of $10^{-5}$). Thus, in the course of simulations A the root-mean-square surface elevation $\sigma$ remains about the same. Significant evolution of neither the peak wavenumber of the average momentary wavenumber spectrum, nor its width was observed.

The spectra of simulated waves in the infinitively deep water, reported in Sergeeva & Slunyaev (2013), correspond to the prescribed initially JONSWAP shape very well. For shallower cases simulated this time the spectra are shown in Fig. 1 by red solid lines. Above the figures the peak and mean cyclic frequencies and wavenumbers are given. The black dashed curves in Fig. 1a,c,e show the JONSWAP frequency spectrum for given $\omega_p$, $\sigma$ and $\gamma$. Fig. 1b,d,f display the corresponding wavenumber spectra; the curves for the JONSWAP wavenumber spectra result from the corresponding JONSWAP frequency spectra taking into account the finite water depth. One may see that the frequency and wavenumber spectra in cases $A_2$ and $A_{1.2}$ agree with the desired JONSWAP shape very well, this conclusion holds for the spectral tails plotted in logarithmic coordinates as well. Case $A_{1.6}$ is not shown here as it is similar to series $A_2$ and $A_{1.2}$.

In the shallowest case $A_{0.8}$ a new long-wave component is excited by the initial condition. These long waves have smaller amplitudes, but may be clearly seen in the wave spectra (Fig. 1e,f). The spectral shape of generated shorter waves is evidently different from the prescribed JONSWAP function. A more detailed examination reveals that the long-wave component corresponds to free waves (i.e., they obey the shallow water dispersion relation); the mean wavenumber of the long waves shifts downwards in the course of evolution. Thus, the situation $A_{0.8}$ is found unsatisfactory for reproduction of the desired spectrum; some extraneous wave dynamics is detected. Therefore, this case $A_{0.8}$ is not considered below.

Interesting, that the frequency spectra for deeper waters (Fig. 1a,c) are much more indented than corresponding wavenumber spectra (Fig. 1b,d). Most likely this peculiarity is caused by the dispersion, which is more significant over deep water and results in different widths of wavenumber and frequency spectra.

The simulations for series E correspond to steeper initial conditions; the waves from time to time reach the breaking onset, what results in numerical instability. For these simulations the order of nonlinearity is reduced to $M = 3$ (four wave interactions are properly resolved) and artificial damping is introduced at small scales in the manner similar to Chalikov (2005). Sometimes wave breaking still occurs; then the wave fields are relaxed with the help of specially designed procedure, which locates the source of the heavy spectral tails and smoothes it out. As a result, in the course of evolution the system loses up to about 10% of its energy (see an example in Fig. 2a). The values of root-mean-square surface elevation $\sigma$ used in Table 1 for series E correspond to the middle values during the 20-min evolution. In the course of the evolution the small-scale part of the spectrum decays faster; the peak wavenumber shifts toward smaller values, and the spectrum somewhat shrinks, see Fig. 3b for case $E_2$. The averaged wavenumber and frequency spectra reproduce the designed JONSWAP spectra only approximately, see Fig. 3. The two effects: spectral downshift and subsidence of the spectral tail may be clearly seen. Case $E_{1.6}$ is not shown in Fig. 3 since it is very similar to the already discussed situations.

## 4. Wave height probability distributions

The exceedence probability distributions for simulated waves at infinite depth (conditions $A_\infty$ and $E_\infty$ from Table 1) were computed in Sergeeva & Slunyaev (2013) by means of space series processing. It is natural to deal with space series, when wave dynamics is integrated in time. However, the space series are noticed to be affected by weak small-scale waves, which may disturb the result of identification of single waves and lead to the difference in analysis of time and space data series.

Here we benefit from having the full spatio-temporal data and use time series of the surface elevation for building the probability distributions. Thus, the approach is fully consistent with the usual conventional processing in oceanography. The simulations result in 100 time series which may be taken at any single location. There are 2048 points along 10 km with available time series, what enriches the statistical data significantly. Since the distance between adjacent locations is less than one wave period, the data from close locations is obviously not fully independent. It is straightforward to realize that this circumstance may result in stepwise function of the exceedence probability in the area of sparse wave height data. The characteristic length of correlation of the closely retrieved wave data was estimated in Sergeeva & Slunyaev (2013) (for space series). It was found to be drastically shorter in nonlinear simulations as compared to the reference linear case. It is also necessary to take into account that the tail of the probability functions is prone to underestimating due to the limited amount of statistical data (see examination of this effect in Kokorina & Pelinovsky (2002)). There are also other effects which may lead to deviation of the height exceedence probability from the theoretical Rayleigh curve, such as broad wave spectrum, nonlinearity and finite depth.

The wave height exceedence probability distributions produced from the simulated data are shown in Fig. 4 for two approaches, zero up- and zero down-crossing excursions in time series (Massel 1996): red solid and blue dash-dotted curves respectively. The horizontal axis is the wave height normalized by $4\sigma$ (which is approximately equal to the significant wave height; see its estimation on the basis of the one-third maximal height $H_{1/3}$ in Table 1).

The simulated probabilities are compared with two theoretical distributions. The classical Rayleigh distribution for narrow banded linear waves over deep water

$$P_R(H) = \exp\left[-\frac{H^2}{8\sigma^2}\right], \qquad (2)$$

is frequently used for the reference. As it is already claimed, it disregards the wave nonlinearity and implies the narrowbandness of the spectrum.

The Glukhovskiy distribution (its classical form is used, see Massel, 1996)

$$P_G(H) = \exp\left[-\frac{\pi}{4\left(1+\frac{n}{\sqrt{2\pi}}\right)}\left(\frac{H}{\overline{H}}\right)^{\frac{2}{1-n}}\right], \quad n = \frac{\overline{H}}{h}, \quad (3)$$

which is a modification of (2), takes into account the finiteness of water depth. Parameter $n$ varies from 0 (infinitively deep water) to 0.5 (surf zone), and for all our simulations does not exceed the value 0.3. The distribution tends to the Rayleigh law when depth $h$ grows to infinity. The Taylor expansion of (3) for small $n$ (deep water) gives

$$P_G(H) \approx \exp\left[-\frac{H^2}{8\sigma^2}\right]\left[1 + n\frac{H^2}{8\sigma^2}\left(\frac{1}{\sqrt{2\pi}} - \ln\frac{H^2}{2\pi\sigma^2}\right)\right], \quad (4)$$

if the deep-water relation $\overline{H} = \sqrt{2\pi}\,\sigma$ is used (which follows from theory for Gaussian statistics). The first exponential term in (4) is just the Rayleigh distribution (2), and the term in parentheses changes its sign when

$$H = \sigma\sqrt{2\pi \exp\frac{1}{\sqrt{2\pi}}} \approx 3.06\sigma. \quad (5)$$

Thus, when the depth decreases and $n$ grows, the probability of small waves with $H < 3\sigma$ increases, while the probability of higher waves decreases.

The following conclusions may be drawn from examination of Fig. 4. In the situation of moderate steepness $A_2$, $A_{1.6}$, $A_{1.2}$ the simulated waves possess probability of high waves about or lower than the Rayleigh distribution; this is even more significant in the shallow cases $A_{1.6}$, $A_{1.2}$. The up- and down-crossing methods do not exhibit noticeable difference. Some rapid decay of the far tails of the functions may be noticed; they are likely caused by the correlation of time series from close locations, discussed above. Bearing this assumption in mind, one may conclude that the probability curves $A_2$ and $A_{1.2}$ exceed the Glukhovskiy distribution; the excursion is larger in the deep case $A_2$. The curve for experiments $A_{1.6}$ (Fig. 4c) is very close to the Glukhovskiy distribution.

When waves are very steep, the difference between the up- and down-crossing processing of time series becomes evident (Fig. 4b,d,f); up-crossing waves have appreciably larger heights. This difference becomes apparent only for waves with heights significantly larger than $8\sigma$, so that all of them belong to the rogue wave population. The probability curves for series E exceed the Glukhovskiy distribution more significantly than for A cases, and for the deeper water cases $E_{1.6}$, and especially $E_2$ they also well exceed the Rayleigh distribution. Meanwhile in the shallow situation $E_{1.2}$ the curves remain below the Rayleigh law (Fig. 4f).

The probability curves for deep water cases $A_2$, $E_2$ do not differ much from the probabilities for the infinitively deep water, obtained on the basis of data from Sergeeva & Slunyaev (2013) (not shown). The Glukhovskiy distribution lies somewhat lower than the Rayleigh distribution even in the infinitively deep water case due to the breakdown of the implied assumption on the relation between $\overline{H}$ and $\sigma$, $\overline{H} = \sqrt{2\pi}\,\sigma$; this circumstance holds also for intermediate depths.

The maximum wave height is about $11.6\sigma$ in the case of $E_2$, it is just slightly less than $11.9\sigma$ attained in simulations of waves over infinite depth, reported in (Sergeeva & Slunyaev, 2013). The maximum wave heights are smaller in the less nonlinear case $A_2$ and in shallower water. The predominance of high up-crossing waves in cases $E_2$, $E_{1.6}$, $E_{1.2}$ corresponds to larger rear wave slopes similar to the rogue wave peculiarity noticed in deep-water 2D simulations by Sergeeva & Slunyaev (2013), and in 3D simulations by Xiao et al (2013). In the in-situ time series recorded near the coast of Brasil de Pinho et al. (2014) have identified 197 up-crossing and 108 down-crossing rogue waves, what gives the ratio of 65% to 35%. The present simulations confirm existence of this asymmetry of high waves in rough sea states in the situations of intermediate depth as well.

On the other hand, for waves of smaller amplitude corresponding to cases A, no significant difference between the numbers of up- and down-crossing rogue waves is observed. This result is in agreement with in-situ observations in the coastal zone of the Baltic Sea under rather mild weather conditions (Didenkulova & Rodin, 2012).

## 5. Typical rogue wave events

The evolution of rogue waves may be considered in more detail, having the spatio-temporal wave data. In particular, the lifetimes of rogue events are obtained, and also the asymmetry of rogue waves at different instants is studied in more detail. The use of spatial periodicity allows to follow the rogue wave during its occurrence and disappearance. Thus, spatial time series are used in this section.

Firstly, the rogue waves which fulfill the condition on the wave height (1) are selected, and the shapes of the surfaces are analyzed. The rogue wave shapes are divided into four classes depending on whether the crest height is larger than the trough height or not (positive or negative waves), and whether the deep trough precedes or follows the high crest (front or rear waves).

The charts characterizing the typical shape of rogue waves are shown in Fig. 5 for two utmost depths; simulations $A_{1.6}$ and $E_{1.6}$ correspond to the intermediate case and are not shown here. Waves with larger crests dominate. Deep-water Stokes waves have higher and sharper crests, thus this observation cannot be surprising in cases $A_2$, $E_2$. Shallow water cnoidal waves are also strongly asymmetric, though cases $A_{1.2}$, $E_{1.2}$, which show even stronger vertical asymmetry than at deeper water, correspond to the intermediate depth $k_p h \sim 1.2$. In the case of steep waves over relatively shallow water $E_{1.2}$ there is practically no rogue waves represented by deep troughs (so-called holes in the sea), see Fig. 5d.

Moderately steep waves do not exhibit asymmetry with respect to the vertical line, though rogue waves in the steeper wave conditions do. The portion of rogue waves with rear slope, which is higher than the front slope, significantly exceeds the fraction of rogue waves with opposite asymmetry in cases E (Fig. 5b, d). These conclusions fully agree with the statements on difference between statistics of up- and down-crossing waves made in the previous section; thus the analyses of time series and space series have resulted in similar findings.

The spatio-temporal data of simulations is stored with high resolution, what allows considering the occurrence of extreme events in detail. When rogue waves selected according to the criterion (1) are detected in consequent space series at close locations, they are consolidated forming clusters of longer rogue events in the time-space wave diagrams. Within the event the condition (1) on the extreme wave may be violated, but only for short times (up to 2.5 wave periods). The eventual lifetimes of the rogue events are plotted in Fig. 6 versus the amplification index, $AI = H/(4\sigma)$ (Fig. 6a) and versus the sequence number of rogue events (Fig. 6b), for two rates of nonlinearity and two water depths.

The number of rogue waves is much more in case $E_2$, which is the deepest and the most nonlinear one. Rogue waves in deep sea seem to amplify stronger (Fig. 6a). Typically, longer living events result in stronger wave amplification. From Fig. 6b one may see some weak regularity that rogue waves which occur in deeper water seem to live longer, though this statement is rather uncertain due to very different number of the events in the cases. There is no significant difference between simulations A and E for the same depths, deep or shallow. It may be noted that the analytic estimation of lifetimes of extreme waves caused by linear dispersive focusing and nonlinear self-modulation, made by Slunyaev & Shrira (2013), resulted in a similar conclusion about proximity of these values. Hence, the effect of nonlinearity seems to contribute insignificantly to the rogue wave life time.

The maximal lifetimes in different shown cases differ about twice, what may be due to a very different number of rogue waves counted in the cases. Most of rogue waves live for a few, up to 10 wave periods; though there is a less numerous population of rogue waves, which may live up to 30-60 wave periods. Thus the upper limit is in agreement with the results reported by Sergeeva & Slunyaev (2013). Of course, the present work does not answer the question about the effect of transverse dimension.

## 5. A case study from the Baltic Sea

A variety of spectral shapes may characterize waves in the coastal area. The JONSWAP spectrum is one of the most used model spectra for wind waves. It has no explicit dependence on water depth, and originally was obtained for relatively deep North Sea. The frequency spectra of waves in coastal waters of Taiwan turned out to fit the JONSWAP function with different values of the peakedness parameter rather frequently (Doong 2015). Using our own database of surface wave records from the Baltic Sea, we examine below if the coastal waves may be described by the JONSWAP spectrum, and what is the corresponding probability function.

The details of the relevant in-situ experiments in the Baltic Sea may be found in Didenkulova & Anderson (2010), Didenkulova (2011), Didenkulova & Rodin (2012). The surface elevation caused by wind waves was recorded at water depth $h = 2.7$ m by ultrasonic echosounder (LOG_aLevel® device) with acquisition frequency 5 Hz. The measurements were performed continuously during rather mild weather conditions in June-July of 2008 and 2009. To separate wind waves from intense ship wake signals observed during the day, we consider only night records for the duration of 7 hours (from 0:00 till 7:00). For our purposes we select randomly a few days with single peaked frequency spectra. Firstly, the selected 7-hour records are cut into 20-min segments; and the samples are investigated searching for periods of quasi equilibrium, characterized by similar-shaped single-peaked spectra with close values of the peak frequency and standard deviation for the surface elevation, $\sigma$. The duration of the obtained quasi equilibrium records is from 2 to 4 hours for different records. Then, frequency spectra are obtained for even shorter wave samples; three examples of the spectra are shown in Fig. 7a (see the legend). The spectra are compared with the JONSWAP function for given $\sigma$ and peak frequency $\omega_p$, and different values of peakedness $\gamma$, from 2 to 6 (gray curves with numbers indicating values of $\gamma$). Amplitudes of the power spectra and frequencies in Fig. 7a are scaled. Due to the difference between the peak frequencies of the records from different days, the effective depth conditions are different; they correspond to $k_p h \approx 1.6...2.6$, where $k_p$ is the peak wavenumber, related to the peak frequency by the linear dispersion law, $\omega_p^2 = g k_p \tanh k_p h$ ($g$ is the gravity acceleration). The characteristic wave steepness defined as $2 k_p \sigma$ varies from 0.10 to 0.13 for these records.

It may be seen from Fig. 7a that the high-frequency tails of the measured waves agree with the JONSWAP spectrum with small $\gamma$ rather well; the agreement is perhaps somewhat worse in the long-wave domain. The most energetic parts of the spectra seem to correspond to the JONSWAP spectrum with $\gamma \approx 2...4$.

Considering the characteristic steepness and depth parameters, the following rough correspondence between the data from the Baltic Sea (see caption for Fig. 7) and the numerical simulations (see Table 1) may be set: the record made on July 1, 2008 is similar to case $A_2$, the record from July 18, 2008 – is similar to $E_{1.6}$, and the time series from 2009 is close to case $E_2$.

The wave height exceedence probability functions for the in-situ time series are plotted in Fig. 7b in a similar to Fig. 4 way (the Glukhovskiy distributions differ for the cases and are not plotted). There are few rogue waves according to condition (1), but the maximal waves exceed the value $4\sigma$ not much. No clear difference between the up- and down-crossing statistics may be found in Fig. 7b. The records from 2008 exhibit wave height probabilities similar to the numerical simulations (Fig. 4a,c); they are mainly below the Rayleigh distribution. The steeper record from 2009 definitively exceeds it similar to Fig. 4b. It may be pointed out that the frequency spectrum of the record from 2009 seems to be more peaked compare to the others. Thus, the considered examples from the Baltic Sea agree with the general picture which follows from the performed numerical study, though the agreement between curves for simulation $E_{1.6}$ and for the record from July 18, 2008 is not excellent.

## 6. Conclusion

Unidirectional surface sea waves with JONSWAP spectrum are simulated numerically in conditions of moderate and strong nonlinearity. The main focus is made on the effect of the finite depth in the intermediate range of $k_p h$ from 2 to 0.8 on the probability and characteristic appearance of rogue waves. Of order of $O(10^7)$ individual waves are used for the statistical analysis in each case. Much attention is paid to the physical adequacy of reproduction of the equilibrium sea state for given conditions. The cases $k_p h \approx 2$, $k_p h \approx 1.6$ and $k_p h \approx 1.2$ are considered in more detail. The findings are compared with our previous work on the infinitively deep water simulations Sergeeva & Slunyaev (2013). The results of numerical simulations are also compared versus the analysis of a few in-situ time series from the coastal zone of the Baltic Sea.

The main outcomes of the study may be formulated as follows.

The spatio-temporal dynamics of the sea waves for given JONSWAP spectrum was successfully replicated in the numerical simulations for conditions $k_p h \approx 2$, $k_p h \approx 1.6$ and $k_p h \approx 1.2$. The both, wavenumber and frequency wave spectra are reproduced accurately; though in the steepest case they are altered by the downshift and decay due to parameterized wave breaking. In the shallowest case $k_p h \approx 0.8$ spurious wave dynamics affects the wave parameters significantly.

The use of time records at different locations helps to increase the amount of statistical data, and to describe rarer events. The wave height exceedence probability functions exceed the Glukhovskiy distribution, but may drop below the Rayleigh law in situations of relatively shallow water and weak nonlinearity. The difference in probabilities between the infinitively deep water condition and the case $k_p h \approx 2$ is rather quantitative; the probability of high waves under intense wave conditions exceeds the Rayleigh distribution.

The maximal wave height achieved in numerical experiments reaches $11.6\sigma$, which is about three times the significant wave height. The rogue events may last for 30-60 wave periods depending on the local depth.

The analyses of time series and space series result in similar conclusions on the typical shape of rogue waves. The waves have strong vertical asymmetry in all situations from deep water to $k_p h \sim 1.2$; the rogue waves are characterized by larger rear slopes when the waves are strongly nonlinear (extreme up-crossing waves are higher/more frequent) in all considered depths $k_p h \geq 1.2$. The horizontal asymmetry becomes apparent for the rare population of sufficiently high waves only.

Randomly chosen in-situ time series from the Baltic Sea, which are characterized by a single-peak steady spectrum, agree with the general picture of the numerical simulations.

The work has received support from RFBR grants 15-35-20563 and 16-55-52019. ASl and ASe gratefully acknowledge support from Volkswagen Foundation. ID thanks Institutional research funding IUT 19-6. The comparative study of the time series processing and the space series processing is performed within the RSF grant No 16-17-00041.

**Table 1.** Parameters of numerical simulations. The shaded columns correspond to the conditions studied in Sergeeva & Slunyaev (2013)

| Series code | $A_{0.8}$ | $A_{1.2}$ | $A_{1.6}$ | $A_2$ | $A_\infty$ | $E_{1.2}$ | $E_{1.6}$ | $E_2$ | $E_\infty$ |
|---|---|---|---|---|---|---|---|---|---|
| $\gamma$ | 3 | 3 | 3 | 3 | 3 | 3.3 | 3.3 | 3.3 | 3.3 |
| $T_p$, s | 10.2 | 10.2 | 9.9 | 9.8 | 9.8 | 10.3 | 10.8 | 10.8 | 10.4 |
| $4\sigma$, m | 3.33 | 3.26 | 3.32 | 3.38 | 3.50 | 6.20 | 6.34 | 6.45 | 6.61 |
| $H_{1/3}$, m | 3.13 | 3.13 | 3.05 | 3.26 | 3.37 | 6.06 | 6.17 | 6.35 | 6.53 |
| $H$, m | 12.4 | 24.8 | 37.3 | 49.7 | $\infty$ | 27.4 | 41.1 | 54.8 | $\infty$ |
| $k_p$, rad/m | 0.062 | 0.049 | 0.043 | 0.041 | 0.042 | 0.044 | 0.038 | 0.036 | 0.031 |
| $k_p h$ | 0.76 | 1.2 | 1.6 | 2.1 | $\infty$ | 1.2 | 1.6 | 2.0 | $\infty$ |
| $2k_p\sigma$ | 0.10 | 0.08 | 0.07 | 0.07 | 0.07 | 0.14 | 0.12 | 0.12 | 0.10 |

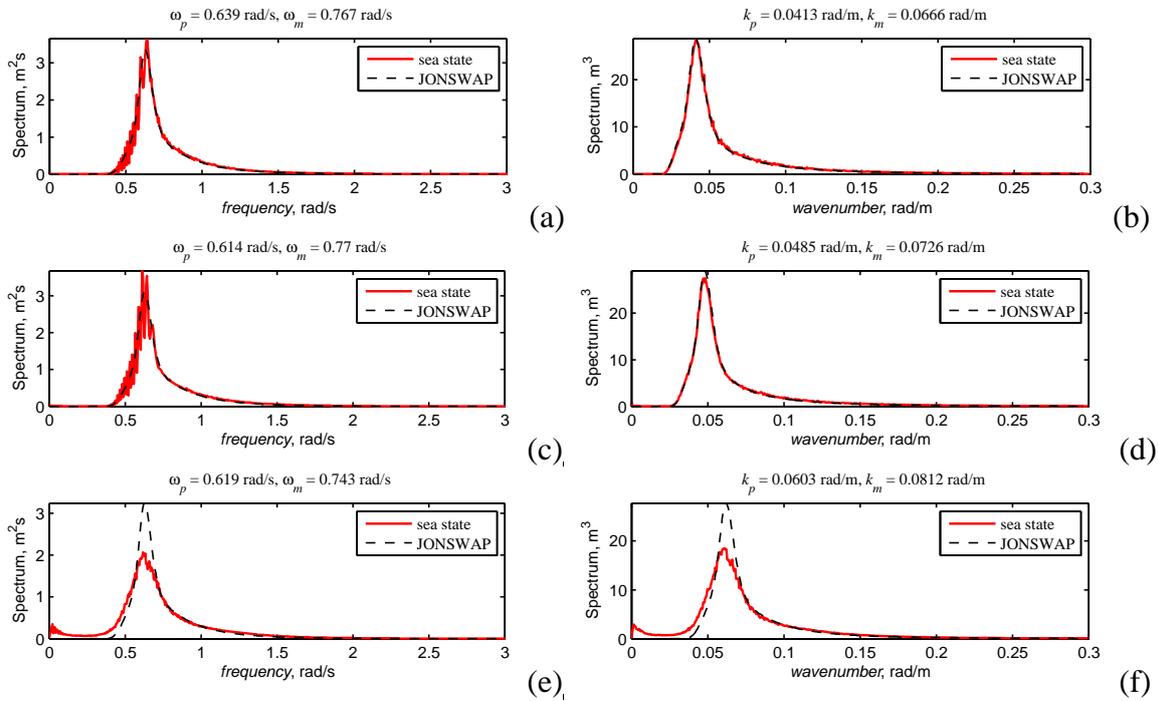

**Fig. 1.** Ensemble averaged frequency and wavenumber spectra for cases $A_2$ (a, b), $A_{1.2}$ (c, d) and $A_{0.8}$ (e, f).

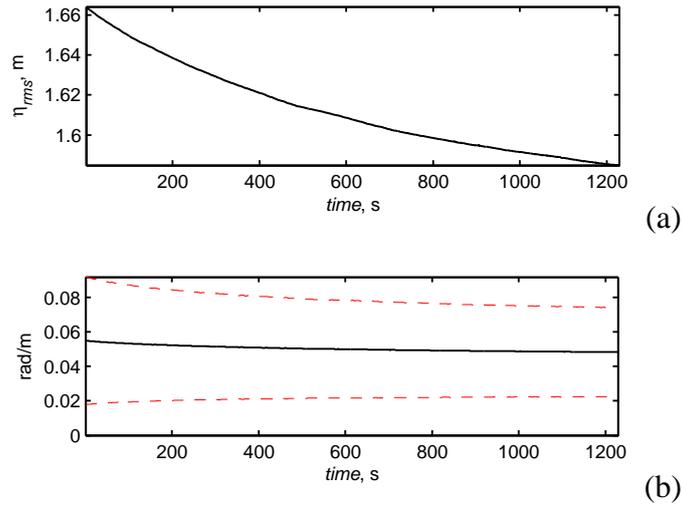

**Fig. 2.** Evolution of the root-mean-square elevation $\sigma$ (a), and evolution of the mean wavenumber with shown characteristic width (estimated through the spectral second momentum) (b); series $E_2$.

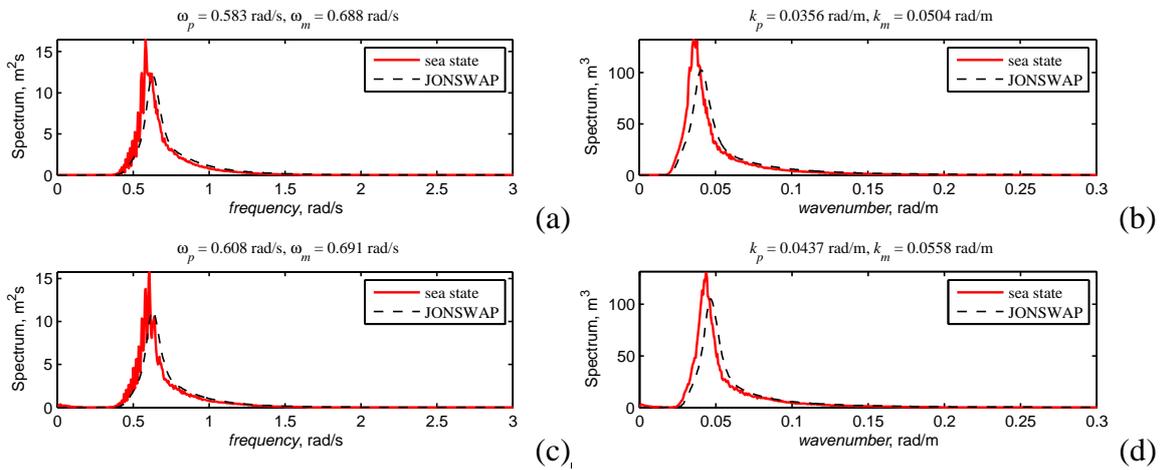

**Fig. 3.** Ensemble averaged frequency and wavenumber spectra for cases $E_2$ (a, b) and $E_{1.2}$ (c, d).

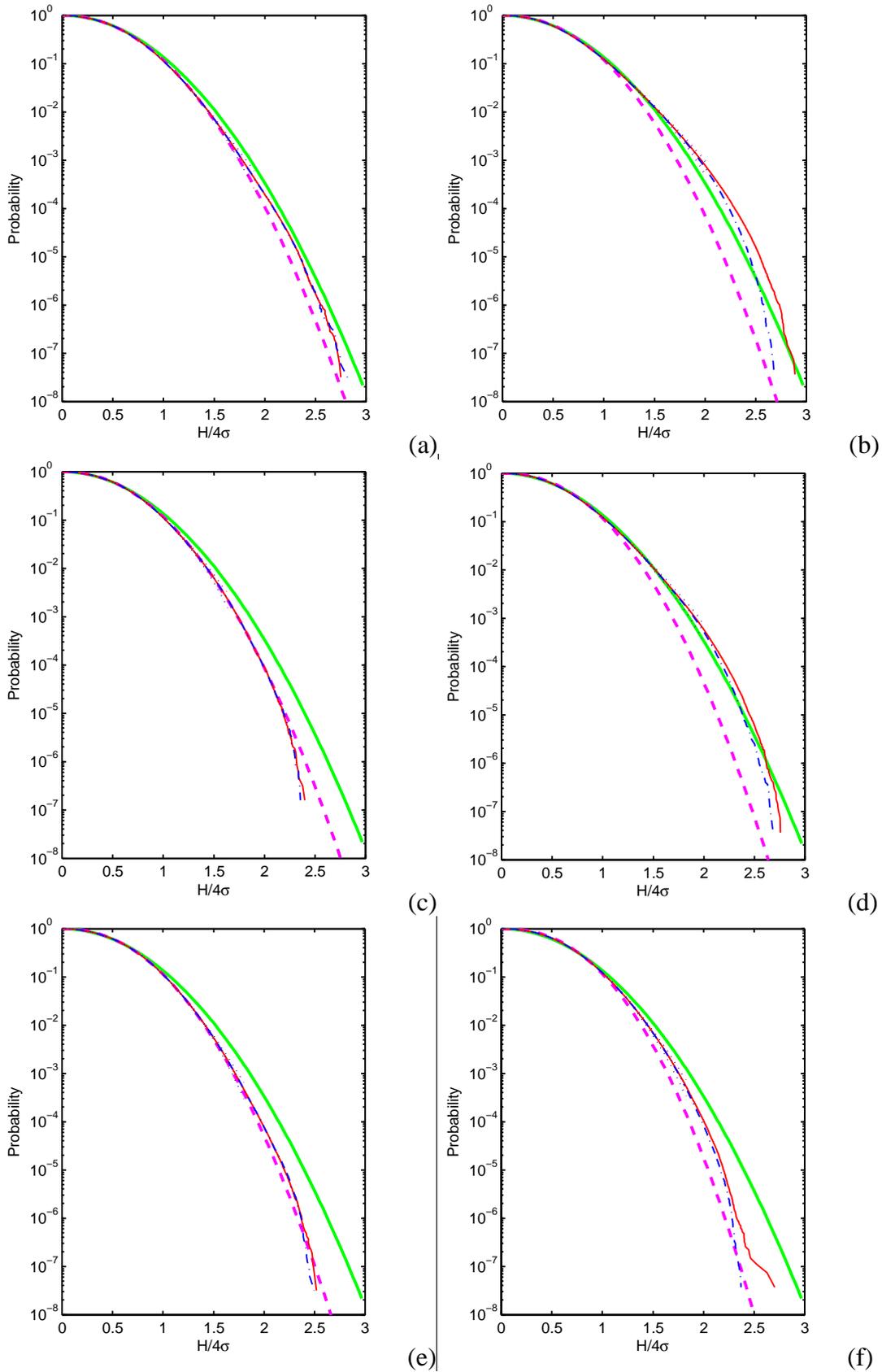

**Fig. 4.** Wave height exceedence probability functions for different cases: $A_2$ (a), $E_2$ (b), $A_{1.6}$ (c), $E_{1.6}$ (d), $A_{1.2}$ (e) and $E_{1.2}$ (f). The thick green solid and magenta dashed lines correspond to the Rayleigh and Glukhovskiy distributions respectively. The thin solid red and dash-dotted blue lines respectively describe results obtained by up- and down-crossing methods.

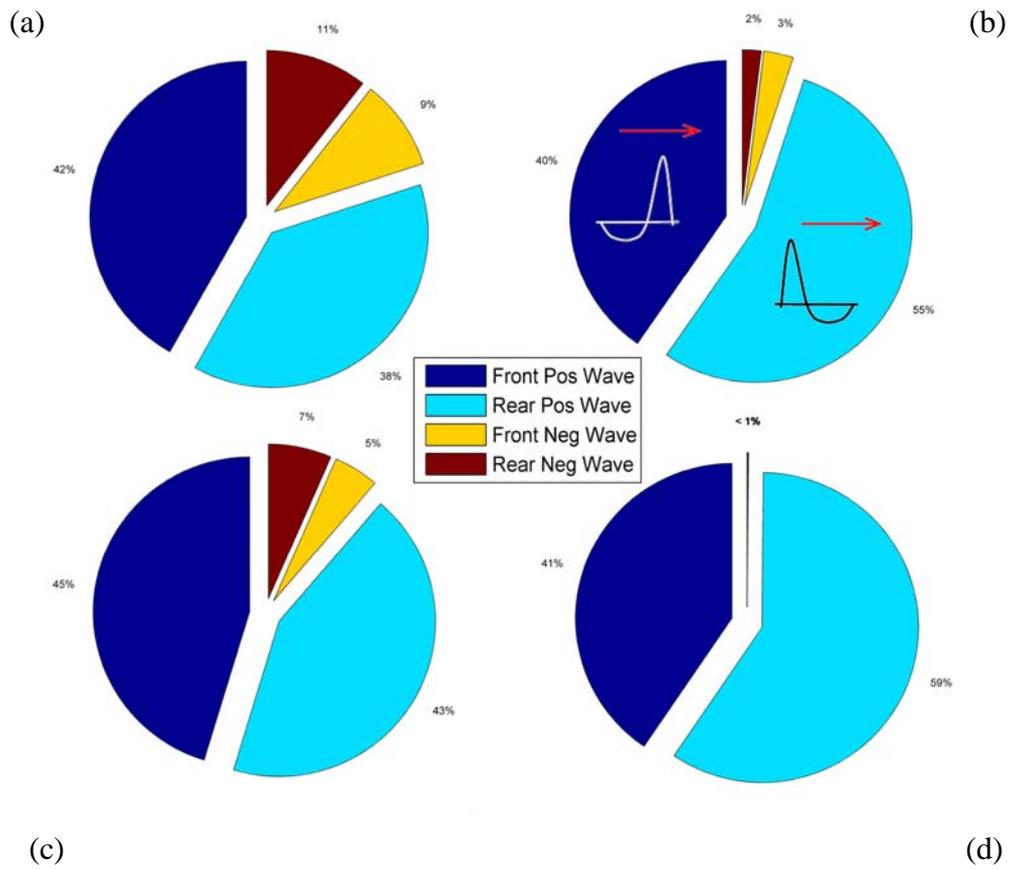

**Fig. 5.** Charts of rogue wave shapes divided in four classes: case $A_2$ (a), $E_2$ (b), $A_{1.2}$ (c), $E_{1.2}$ (d). The characteristic wave shapes are given in (b), where arrows indicate the direction of wave propagation.

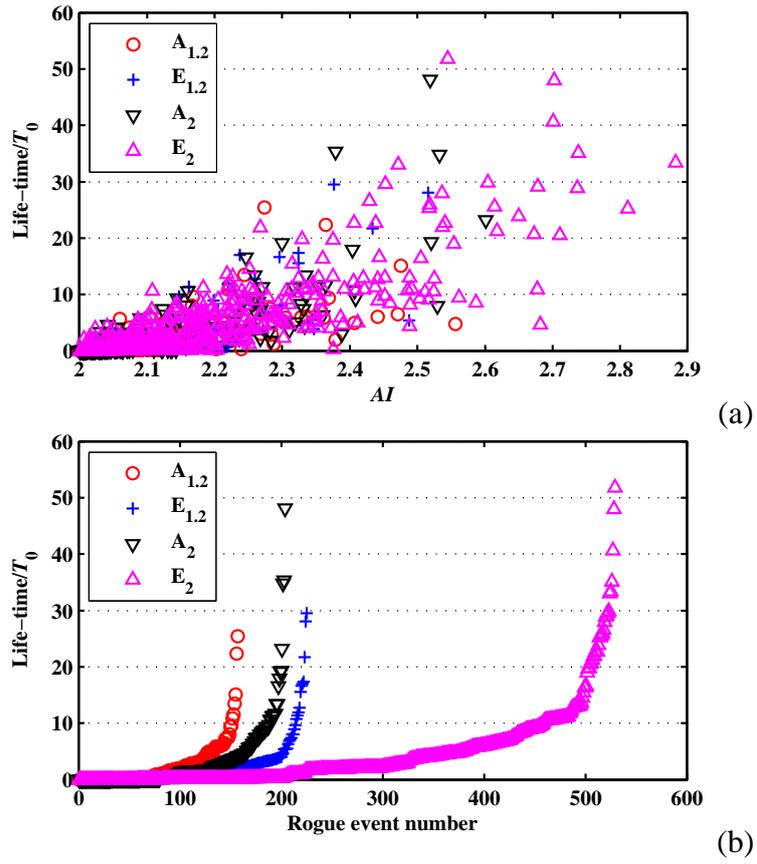

**Fig. 6.** Rogue event lifetimes versus wave amplification (a) and sorted in ascending order (b).

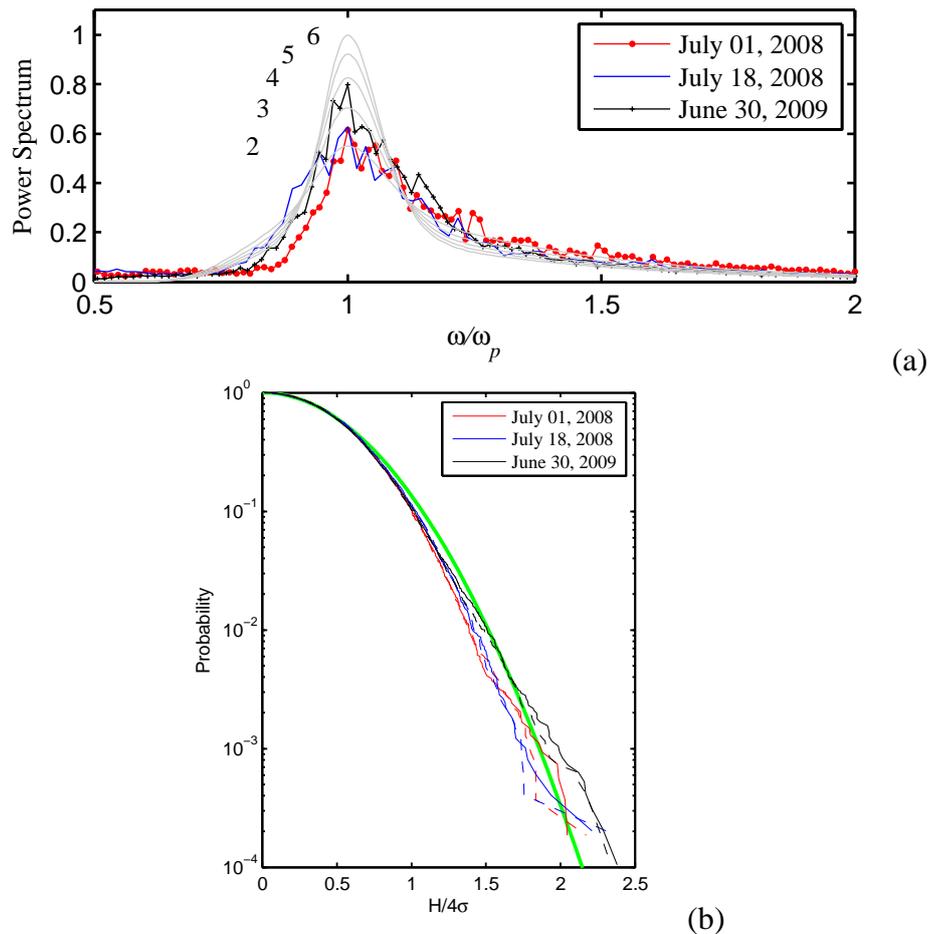

**Fig. 7.** (a): Examples of in-situ wave spectra from the Baltic Sea versus the JONSWAP spectrum. The numbers on the left of the curves denote values of the spectrum peakedness $\gamma$. (b): The corresponding wave height exceedence probabilities. Solid and broken curves correspond to up- and down-crossing wave analysis respectively; the wide green curve reproduces the Rayleigh law. The examples in (a), (b) correspond to dimensionless depths $k_p h \approx 2.6, 1.6, 2.5$ and steepnesses $2k_p \sigma \approx 0.10, 0.13, 0.13$ correspondingly for the consequent dates (see the legends).